# 3DES ECB Optimized for Massively Parallel CUDA GPU Architecture

Lukasz Swierczewski

Computer Science and Automation Institute
College of Computer Science and Business Administration in Łomża
Lomza, Poland
luk.swierczewski@gmail.com

*Abstract*— **Modern computers have graphics cards with much higher theoretical efficiency than conventional CPU. The paper presents application possibilities GPU CUDA acceleration for encryption of data using the new architecture tailored to the 3DES algorithm, characterized by increased security compared to the normal DES. The algorithm used in ECB mode (Electronic Codebook), in which 64-bit data blocks are encrypted independently by stream processors (CUDA cores).**

*Keywords-DES, cryptography, parallel algorithms, CUDA, GPU computing*

## I. INTRODUCTION

Powerful GPUs support CUDA technology for several years are available at most home computers. Computational capabilities of modern GPUs are much higher than traditional CPUs. A typical home has a graphics card performance of 600 GFLOPS (e.g. nVidia GeForce GTS 450). However, there are solutions such as the GeForce GTX 590, which have a capacity up to 2506 GFLOPS. For comparison, a fast six-core Intel Core i7 980 XE has a capacity of only 109 GFLOPS.

The research centers are usually applied card with nVidia Tesla family. They are created with the aim of carrying out scientific or engineering calculations and revised the GeForce cards are widely available. The main difference between the GeForce and Tesla cards is also a performance in carrying out calculations on floating point double precision. The GeForce are largely limited in this respect, which often are not suitable for conducting scientific calculations but are very useful for example in computer games.

The market can also find competing solutions for AMD's Radeon names (home use) and FireStream (use of professional computing). They also have great potential as chips from nVidia, but not so well accepted and are a little less popular.

When the graphics cards enable virtually any calculations appeared in the home computers available to the users get a new solution. It allows the realization of such encryption and decryption in real-time data on the disk. The new massively parallel multi-core GPU can in many ways to replace the CPU.

This paper will present the possibility of parallelization 3DES CUDA-enabled. Triple DES with three different keys (3TDES) has a strength of 168 bits (including parity bit 192 bits). Although the attack in the middle meet the force drops to $2^{112}$ is a cipher to date have not managed to break.

## II. ALGHORITM

### A. Overview Triple DES

Otherwise known as 3DES Triple DES was established in the late 90s because of the inadequate security provided by the Data Encryption Standard. In July 1998, the EFF DES cracker machine built by the Electronic Frontier Foundation broke a DES key in 56 hours. Therefore there is need for more effective solutions than normal DES. The proposed uses 3DES triple DES encryption so that it is estimated that the time needed to break this algorithm is about a thousand years. Currently, a frequently used algorithm is AES (Advanced Encryption Standard), which is the official successor to DES. It has a physical support for new processors Intel AVX instruction form.

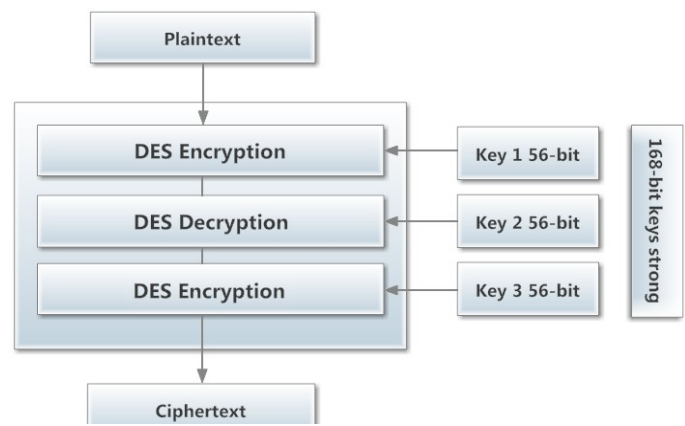

Figure 1. The process of data encryption 3DES.

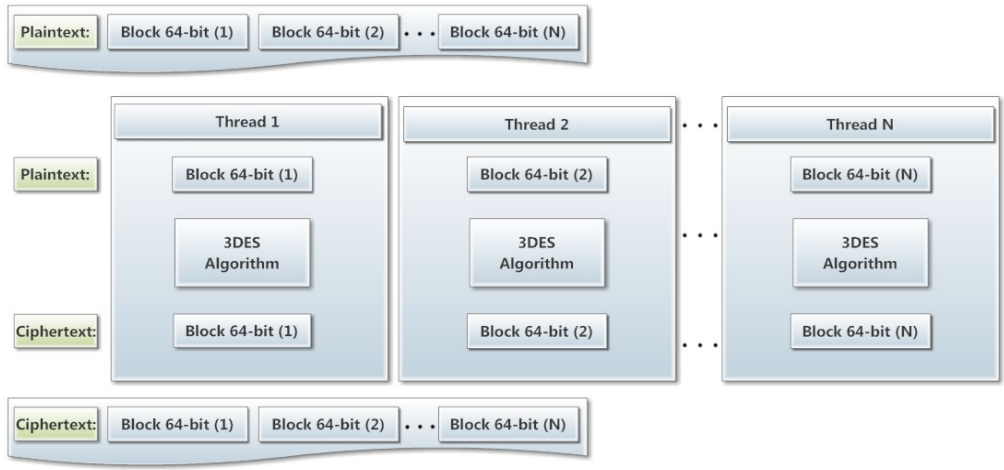

Figure 2. Functional diagram of the 3DES ECB mode.

In the 3DES algorithm, the data are:

a) *encrypted using the first key*,
b) *decrypted using the second key*,
c) *encrypted using a third key*.

This process shows the Fig. 1.

Computational complexity of 3DES is three times higher then the standard DES algorithm. 3DES is a much safer place. Decryption procedure is performed identically only in reverse.

Listing 1.1. Call the kernel.

```
cudaMemcpy(data_device, data_host, size_data * sizeof(long long int), cudaMemcpyHostToDevice);
cudaMemcpy(keys_device, keys_host, 48 * sizeof(long long int), cudaMemcpyHostToDevice);

unsigned long int N = 131072;

unsigned int block_size = 256;
unsigned int n_blocks = N/block_size+(N%block_size == 0 ? 0:1);

unsigned long long int i;
for(i = 0; i <= size_data; i += N)
{
    kernel_3des <<< n_blocks, block_size >>>
        (data_device+i, N, keys_device);
}
cudaMemcpy(data_host, data_device, size_data * sizeof(long long int), cudaMemcpyDeviceToHost);
```

B. *Parallel GPU Algorithm*

Single kernel executes the GPU is processing 131 072 blocks of 64-bit. This gives 1 048 576 bytes or 1 MB. Increasing the number of operations per kernel does not increase performance in any way. This can be seen in Table 2. For maximum performance, all data is copied once to the graphics card, and after completion of all required calculations copied to RAM.

Module of code responsible for calling the kernel is presented in Listing 1.1.

III. RESULTS

Algorithm performance was tested on three processors:

a) *Intel Core 2 Quad Q8200*,
b) *Intel Core i7 950*,
c) *Intel Xeon E7 - 4860*.

and two graphics cards:

a) *nVidia GeForce GTS 250*,
b) *nVidia Tesla C2050*.

Given that the CPUs have been taken into account and given the effects of the algorithm, allowed using variable amounts of threads (cores), based utilizing the library OpenMP. A thorough analysis should be able to draw better conclusions when comparing the CPU with the GPU. A thorough comparison of the specifications of equipment used are presented in Table 3. Presented there, the column "*Performance*" refers to the performance achieved in the case of single precision floating point numbers.

The conducted experiments show that even the slowest included GeForce GTS 250 is faster than one of the fastest processors available today Xeon E7-4860 (ten cores, twenty threads). Even the platform was built with four Xeon processors E7-4860, the total cost at the date of writing is about 15 352 dollars gets worse results than the GeForce GTS 250 costing around 100 dollars (video card is almost two times faster). Also, operating costs appear to be much lower. Four Xeon processors consume up to 520 watts, which is the value

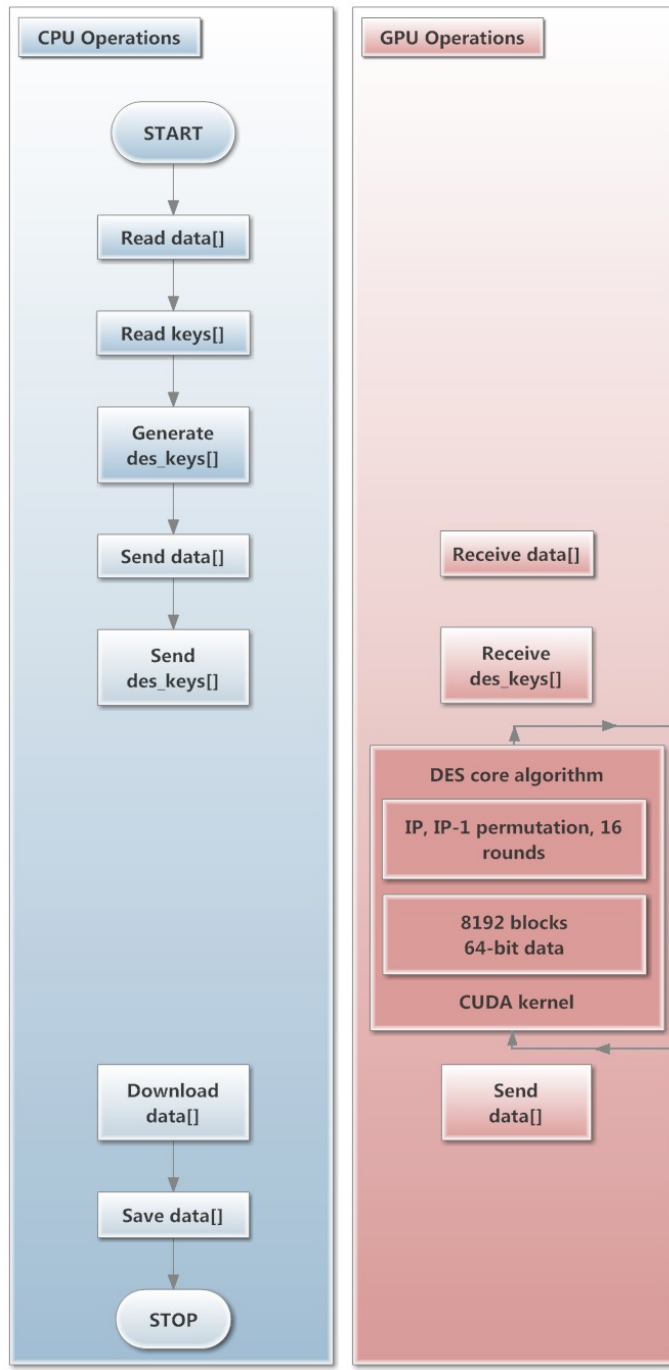

Figure 3. Schematic distribution of computation between the CPU and GPU.

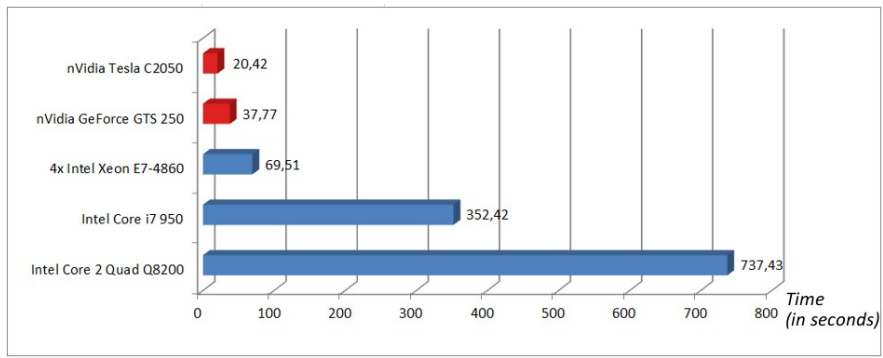

Figure 4. Time data encryption 3DES ECB on different processors.

to 3.46 times higher then the demand for electricity by the GTS 250 (only 150 watts).

Summarize both the purchase of equipment and its maintenance is a very strong point of graphics accelerators. Additionally, having an appropriate motherboard, you can easily add an additional graphics card, which could theoretically increase the utilization of double productivity. But we must remember that, then the developer will be saddled with the additional task of the division of tasks between the graphics cards.

There was quite a significant difference in performance between the GTS 250 system and Tesla C2050 which is a natural thing. Tesla returns the results of 1.84 times faster than GeForce. Although the GTS 250 has only 128 SP and memory bandwidth of 70.4 GB/s which is very low compared to the SP 448 and 144 GB/s shared by Tesla-class solution is due to higher SP clock frequency (1836 MHz in the case of the GTS 250, and 1150 MHz for the Tesla C2050) theoretical maximum efficiency so does not significantly differ. C2050 processor GFLOPS rate is 1050, and for up to 705 GFLOPS GeForce. It should be remembered that the GTS 250 card is the older generation and is based on the engine Unified Shaders, which are less able than nVidia FERMI.

TABLE I. DEPENDENCE OF THE SIZE OF A DEFINED BLOCK OF RUN-TIME KERNEL (FOR GEFORCE GTS 250).

| Block size: | 8 | 16 | 32 | 64 |
|---|---|---|---|---|
| Execution time *(in seconds)*: | 96.184 | 45.844 | 22.985 | 20.452 |
| Block size: | 128 | 256 | 512 | |
| Execution time *(in seconds)*: | 20.676 | 20.424 | 21.186 | |

TABLE II. DEPENDENCE OF THE NUMBER OF OPERATIONS PERFORMED BY A SINGLE KERNEL OF THE TOTAL EXECUTION TIME (FOR TESLA C2050). KERNEL SIZE DEFINED AS THE NUMBER OF BLOCKS 64-BIT.

| Block size: | 128 | 256 | 512 | 1024 | 2048 | 4096 |
|---|---|---|---|---|---|---|
| Execution time *(in seconds)*: | 536.34 | 310.2 | 155 | 77.7 | 38.78 | 35.7 |
| Block size: | 16384 | 32768 | 65536 | 131072 | 8388608 | |
| Execution time *(in seconds)*: | 22.43 | 22.12 | 21.04 | 20.42 | 22.68 | |

In the presented results (Table 4, Fig. 4) includes only the total execution time of all the kernels. Was not taken into account the time required to perform input, output and data transfer between RAM and graphics card. Time spent on these additional operations was always around 10.404 seconds for an encrypted file size of 512 MB. Most of the time it takes to perform operations that require the involvement of the hard drive. Exchange of data on the PCI-E v2.0 is possible with a bandwidth of 8 GB/s (full duplex).

TABLE III. TECHNICAL DESCRIPTION OF THE PROCESSORS USED IN COMPUTING.

| Name | Performance | TDP | Ratio TDP/ GFLOPS |
|---|---|---|---|
| Intel Core 2 Quad Q8200 | 37.00 GFLOPS | 95W | 2.56 |
| Intel Xeon E7- 4860 | 90.64 GFLOPS | 130W | 1.43 |
| Intel Core i7 950 | 49.06 GFLOPS | 130W | 2.65 |
| nVidia GeForce GTS 250 | 705 GFLOPS | 150W | 0.21 |
| nVidia Tesla C2050 | 1050 GFLOPS | 238W | 0.22 |

IV. CONCLUSION

With GPU acceleration can be significant. Even less of the available graphics cards are much faster than the CPU. The 3DES algorithm used, however, restricted mainly to the bit operation. In addition, are used for integer operations, where the GPU achieve great results. In the case of algorithms based on the variables of double precision floating-point difference between the GPU and the CPU would still be very visible.

Disadvantage in some applications graphics cards may be available memory. The top model nVidia Tesla C2070 has 6 GB of RAM. This is a fairly small amount compared to the computer's memory, which is usually in the more advanced workstations is at least 24 GB.

Also very interesting issue may be able to use OpenCL environment, which is an intermediate layer between the programmer and the hardware. With OpenCL one code can be compatible with both nVidia graphics cards (GeForce, Tesla), AMD (Radeon, FireStream) and applied eg CELL processor IBM Blade units and PlayStation 3 consoles. Such a range of solutions is of great potential and prospects for the study of parallel algorithms.

TABLE IV. THE RESULTS OF THE CPU AND GPU.

| CPU / GPU name | Frequency | Threads (cores)* | Speedup relative to: | | | | Time (in seconds) |
|---|---|---|---|---|---|---|---|
| | | | Single CPU thread | Intel Core 2 Quad Q8200** | Intel Core i7 950** | Intel Xeon E7 - 4860*** | |
| Intel Core 2 Quad Q8200 | 2,33 GHz | 1 (1) | 1 | - | 0,12 | 0,02 | 2905,22 |
| | | 2 (2) | 1,98 | - | 0,24 | 0,05 | 1464,66 |
| | | 4 (4) | | - | 0,48 | 0,09 | 737,43 |
| Intel Core i7 950 | 3,07 GHz | 1 | 1 | 0,34 | - | 0,03 | 2177,23 |
| | | 2 | 1,99 | 0,67 | - | 0,06 | 1093,86 |
| | | 4 | 3,99 | 1,35 | - | 0,12 | 544,43 |
| | | 8 (4) | 6,17 | 2,09 | - | 0,19 | 352,42 |
| Intel Xeon E7- 4860 | 2,27 GHz | 1 | 1 | 0,24 | 0,11 | - | 3115,95 |
| | | 80 (40) (4x CPU) | 44,82 | 10,61 | 5,07 | - | 69,51 |
| **nVidia GeForce GTS 250** | **1,86 GHz** | **128 CUDA cores** | - | **19,52** | **9,33** | **1,84** | **37,77** |
| **nVidia Tesla C2050** | **1,15 GHz** | **448 CUDA cores** | - | **36,11** | **17,26** | **3,4** | **20,42** |

\* number of cores is optional; ** all the processor cores; *** four processors

ACKNOWLEDGMENT

The work has been prepared using the supercomputer resources provided by the Faculty of Mathematics, Physics and Computer Science of the Maria Curie-Skłodowska University in Lublin.